# Machine Learning-based Regional Cooling Demand Prediction with Optimized Dataset Partitioning

**Meng Zhang[1], Zhihui Li[2*], Zhibin Yu[3*]**

[1]*Mechanical and Aerospace Engineering, School of Engineering, University of Liverpool, Liverpool, L69 7ZX, UK*
[2*]*Mechanical and Aerospace Engineering, School of Engineering, University of Liverpool, Liverpool, L69 7ZX, UK*
[3*] *Mechanical and Aerospace Engineering, School of Engineering, University of Liverpool, Liverpool, L69 7ZX, UK*

**Abstract**

In the context of global warming, even relatively cooler countries like the UK are experiencing a rise in cooling demand, particularly in southern regions such as London. This growing demand, especially during the summer months, presents significant challenges for energy management systems. Accurately predicting cooling demand in urban domestic buildings is essential for maintaining energy efficiency. This study introduces a generalised framework for developing high-resolution Long Short-Term Memory (LSTM) and Gated Recurrent Unit (GRU) networks using physical model-based summer cooling demand data. To maximise the predictive capability and generalisation ability of the models under limited data scenarios, four distinct data partitioning strategies were implemented, including the extrapolation, month-based interpolation, global interpolation, and day-based interpolation. Bayesian Optimisation (BO) was then applied to fine-tune the hyper-parameters, substantially improving the framework predictive accuracy. Results show that the day-based interpolation GRU model demonstrated the best performance due to its ability to retain both the data randomness and the time sequence continuity characteristics. This optimal model achieves a Root Mean Squared Error (RMSE) of 2.22%, a Mean Absolute Error (MAE) of 0.87%, and a coefficient of determination ($R^2$) of 0.9386 on the test set. The generalisation ability of this framework was further evaluated by forecasting cooling demand for the year 2050, demonstrating its robustness in out-of-domain prediction tasks.

## 1. Introduction

Global warming has intensified extreme weather events, causing significant impacts on the environment, economy, and human life. According to the World Meteorological Organization (WMO), there is a 66% probability that global temperatures will exceed 1.5°C above pre-industrial levels between 2023 and 2027, and the next five years may be the hottest on record [1]. With rising global temperatures and changing climate patterns, the growth in cooling demand in the UK is inevitable and requires further in-depth investigation in future research. A study found that air conditioning accounts for around 10% of the total electricity demand in the UK, with an estimated annual growth of 2% [2]. By 2060, global energy demand for cooling is expected to equal that of heating [3]. Given the increasing demand for cooling in domestic buildings, accurately forecasting cooling demand is essential for optimising energy usage and ensuring the stability of energy grids. However, one of the key challenges in predicting cooling demand is the lack of detailed historical data, particularly for domestic buildings in the UK, such as those in London. Moreover, large-scale cooling demand models often focus on aggregate predictions rather than high-resolution hourly forecasts. This scarcity of data makes it difficult to develop robust and high-resolution predictive models. To support effective energy management and promote sustainable urban development, accurate forecasting of cooling demand has become increasingly essential. Many studies have focused on developing models that can reliably capture cooling demand patterns. Currently, cooling demand prediction primarily employs three methodologies: economic data-based models, physical modelling and data-driven models. These approaches provide a strong foundation for understanding cooling demand patterns, each with its own strengths and limitations.

Economic data-based models are often employed by economic or governmental sectors to estimate the annual cooling demand of a region. This approach rely on economic and demographic indicators, are therefore efficient and computationally less demanding, making them well-suited for large-scale projections. Economic data-based models are used for projecting future cooling needs by combining socio-economic factors with climate indicators such as Cooling Degree Days (CDD) [4]. These models leverage the strong correlation between CDD and cooling energy demand, along with economic factors such as population, energy consumption, and GDP, to provide region-specific predictions. Similar cooling demand model were developed based on socio-economic and climatic inputs, centering on the CDD method [5], [6], [7]. The correlation between cooling demand and meteorological conditions was examined in the US and applied these findings to estimate European domestic cooling demand [8], [9]. Additionally, research in HK and UK further illustrate CDD's versatility in predicting cooling demand across different regions and scenarios [10] ,[11]. Despite their widespread use, economic data-based models have notable limitations. Their precision is generally insufficient for detailed energy demand analyses, as they are primarily designed for high-level economic assessments. These models often struggle to capture time-sensitive demand patterns, making them unsuitable for generating detailed temporal profiles of cooling demand. As a result, their utility is confined to broader regional or annual estimates, rather than fine-grained, time-resolved forecasting.

Physical models, grounded in heat balance principles, offer highly accurate and dynamic predictions by incorporating detailed building and environmental parameters. Such models utilize parameters like external and internal temperatures, thermal resistance, and building envelope properties to simulate the flow of heat between a building's interior and its surroundings. Koo et al. developed a mathematical model that can be used for building cooling and heating requirements using Lagrangian finite element method to simplify the physical model-based building simulation approach to a mathematical model [12], [13]. Most studies have shown that physical models can be effectively applied to both regional and building-specific energy prediction with different scenarios [14], [15], [16]. Another common area of research focuses on analysing the impact of various parameters on physical models to improve their accuracy[17], [18]. These studies consider factors such as occupants[19], regional diversity [20], types of weather data [21], urban factors [22] thermal inertia [23], window-to-wall ratios and integrating detailed environmental interactions [24]. While these physical models and simulation tools offer precise assessments of energy demand, they often require substantial computing power and specific building data. These constraints highlight the need for scalable alternatives that address the limitations in data and resources, particularly for large-scale urban studies. In data-scarce areas like most parts of the UK, this method faces significant limitations.

While the aforementioned physical models provide detailed insights into cooling demand, their reliance on extensive environment data and computational resources has driven interest in data-driven approaches. Data-driven models rely on historical cooling demand data to establish relationships between inputs and outputs, making them highly effective in regions with available data. Machine learning is a key component of data-driven methods, playing a crucial role especially in prediction tasks. Regression is one of the most widely used methods for predicting cooling demand due to its simplicity and effectiveness in modelling the relationship between various influencing factors and cooling demand [25], [26], [27]. In addition, there are regression-based models that have been further improved [28]. Techniques such as artificial neural networks (ANN) and support vector machines (SVM), random forest are commonly used to analyse large datasets, identifying complex, nonlinear relationships between input features (e.g., weather, occupancy) and energy demand in buildings [29], [30], [31]. Among them, ANN, as an efficient and straightforward model, is widely used in energy forecasting [32], [33]. LSTM outperforms ANN in handling sequential data by capturing long-term dependencies through its memory cells while also mitigating the vanishing gradient problem, making it more effective for time-series forecasting and complex temporal relationships [34], [35]. These models are highly flexible and capable of producing high-resolution predictions, making them particularly suitable for dynamic and nonlinear systems. However, their performance heavily relies on the availability of extensive and high-quality training data, as insufficient or biased data can significantly degrade model accuracy.

A practical challenge to address is how to develop a high-resolution and fast model for large-scale applications. This study develops a general methodology that employs a physical model to generate cooling demand data for domestic buildings in large cities, enabling the training of high-resolution data-driven models for urban cooling demand prediction. As far as the authors are concerned, this is the first attempt to combine physical models and machine learning models to predict the hourly cooling demand of a megacity. The methodology of the model includes building simulation, results validation, model training, and optimisation, which are introduced in Section 2. The results from different machine learning models are then compared and discussed in Section 3. The key conclusions are then drawn in Section 4.

## 2. Methodology

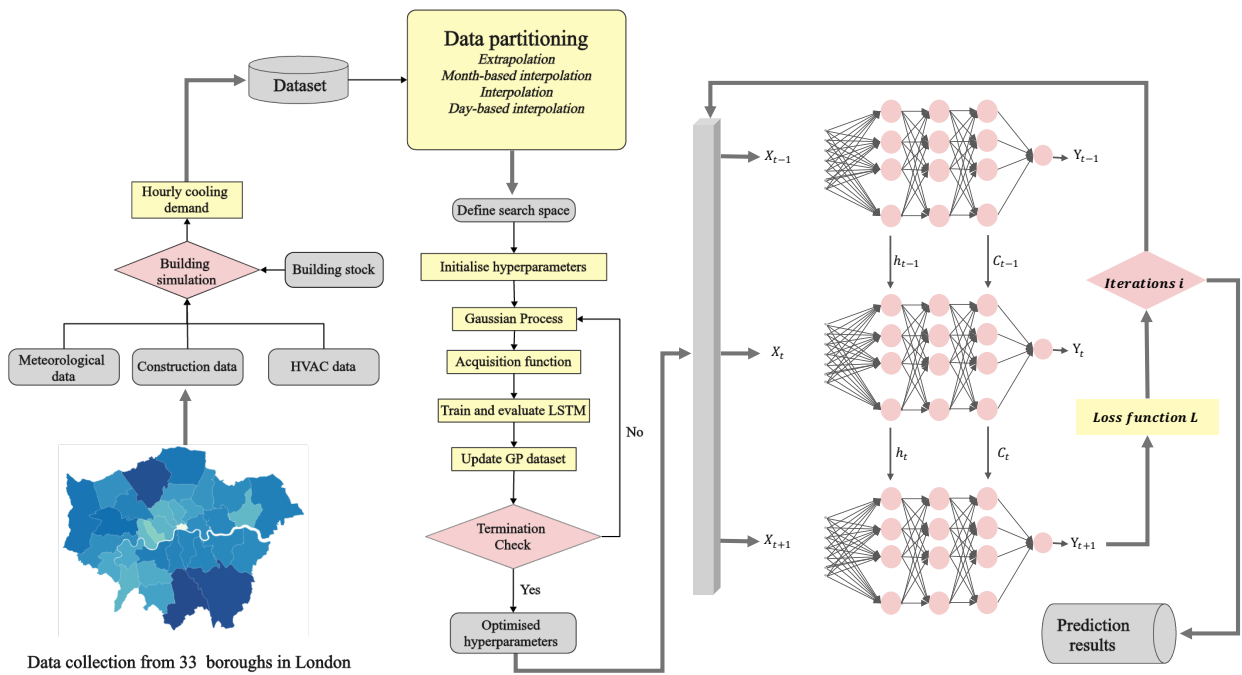

**Figure 1 Block diagram of the model**

Figure 1 illustrates the overall modelling process. The physical model mainly consists of building simulations and building stock data. Meteorological data for the London area is historical records from Solcast [36], and the cooling demand data generated by the numerical simulations, are used as training data to develop LSTM and GRU networks. Additionally, BO is applied to refine the model structure and parameters, thereby enhancing predictive accuracy. After confirming the effectiveness of the optimisation, extrapolation, month-based interpolation, interpolation, and day-based interpolation are compared to evaluate the model's generalisation ability under different train-test split strategies. A detailed description of this process is provided in the following section.

### 2.1 Building simulation

Building simulation for cooling demand prediction primarily rely on three key inputs. The first input is building stock, which determines building types and insulation levels. The second input is weather data, which defines the external environmental conditions that the building is exposed to. The third input consists of building structural data, such as U-values and material properties, which influence heat transfer rates and thus affect cooling requirements. As is shown in Figure 2. The physical model is built using

*DesignBuilder* software, which integrates *EnergyPlus* as its calculation engine. The study focuses on residential buildings in London, divided into four primary types: detached, semi-detached, terraced, and flats. London is further divided into 33 boroughs, each with unique building characteristics and insulation levels. Based on these variations, buildings are segmented by three distinct insulation levels that align with construction periods: pre-1919, 1919-1964, and post-1964. This classification generates 12 unique cooling profiles per borough, resulting in a total of 396 profiles that offer a comprehensive overview of London's cooling needs across various building types and regions.

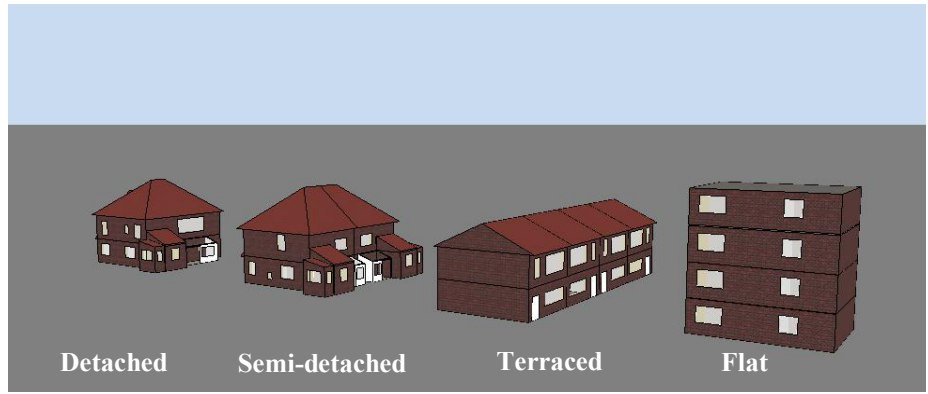

**Figure 2 Models of domestic buildings in *DesignBuilder***

Meteorological data is a key input for these simulations, as it defines the climatic conditions that drive cooling demand. Actual weather data from 2020 is utilized, providing a precise basis for present-day demand estimation. This weather data covers essential variables such as dry-bulb and wet-bulb temperatures, atmospheric pressure, relative humidity, wind speed, wind direction, and various solar irradiance metrics (e.g., global horizontal irradiance). By including these factors, the model captures present weather conditions, offering insights into how climate affects cooling demand. In terms of building characteristics, the model assigns U-values to different building components, such as external walls, roofs, glazing, and ground floors. These U-values reflect heat transfer rates, with each value adjusted based on the building's age and construction period.

The study standardizes internal conditions for consistency across simulations. Occupancy density is set at 0.0196 people per square meter, and a simplified HVAC (Heating, Ventilation, and Air Conditioning) model is used to reflect common cooling practices. The cooling system activates when indoor temperatures exceed 28°C, aiming to cool spaces to a setpoint of 24°C. This controlled setup represents typical daytime cooling needs, providing a baseline estimate of demand. Ventilation is assumed to be idealized across all buildings, incorporating both natural and mechanical systems. While real-world constraints such as building design limitations or safety concerns may limit ventilation in certain cases, this assumption provides a conservative estimate of cooling demand under optimal ventilation conditions. It represents the minimum cooling requirement for London, given the specified climate inputs and structural assumptions.

In addition to building structure, weather, and set point temperature, the cooling schedule is another critical parameter that determines cooling demand, especially the shape of the profile. Different cooling schedules are set for various building types or rooms to simulate human activities within the space. For example, for office workers, the cooling system is set to "on" during weekdays from 5 a.m. to 10 a.m. to align with wake-up times, then switched off until 6 p.m. when it turns on again. For weekends or holidays, the cooling system for the living room operates from 8 a.m. to 6 p.m., while the bedroom cooling is activated during nighttime hours. The timing may vary for different buildings, but the overall schedule follows this general rule. Similarly, the interaction of building characteristics, weather, temperature settings, and cooling schedules generates varying demands. When these demands are aggregated, they represent London's overall cooling demand. As a result, these profiles do not entirely correspond to a single specific heating schedule, as each day exhibits distinct variations.

2.2 Model validation for building simulation

Validating the proposed model for cooling demand forecasting presents unique challenges due to the limited availability of domestic cooling demand data in the UK. Unlike heating demand validation, which relies on comparing total annual demand against publicly available government data, cooling demand validation requires an alternative approach. This study addresses this gap by utilizing benchmarks provided by the London Government, specifically focusing on average peak month cooling demand per square metre. The London Government offers two benchmark methods to guide cooling demand estimation [37]. The first method, rooted in the Standard Assessment Procedure (SAP), is a national framework for calculating residential cooling demand. While SAP provides a consistent and reliable reference, it adopts a static approach that primarily accounts for temperature variations, neglecting other influential factors. The second method, a newly proposed benchmark (PB), refines these limitations by tailoring the calculations to London-specific scenarios. This dynamic approach considers variations in building energy balance and cooling requirements under different conditions, making it a more flexible and comprehensive validation tool.

The cooling demand thresholds for flats and duplexes, as defined by the SAP and BP standards. In this study, building types such as detached, semi-detached, and terraced houses are grouped under the duplex category to align with the broader framework used for comparison. These benchmarks categorize buildings into two distinct groups based on external conditions, one representing optimal conditions and the other reflecting extreme conditions, thereby establishing the range of cooling demand under these scenarios. The average cooling demand values for all building types, representing the average monthly cooling demand per square meter for June, July, and August, are illustrated in Figure 3 using a bar chart, with red and green ranges representing the SAP and BP standards, respectively, spanning from poor to good. The comparison clearly shows that all average values fall within the acceptable standard ranges.

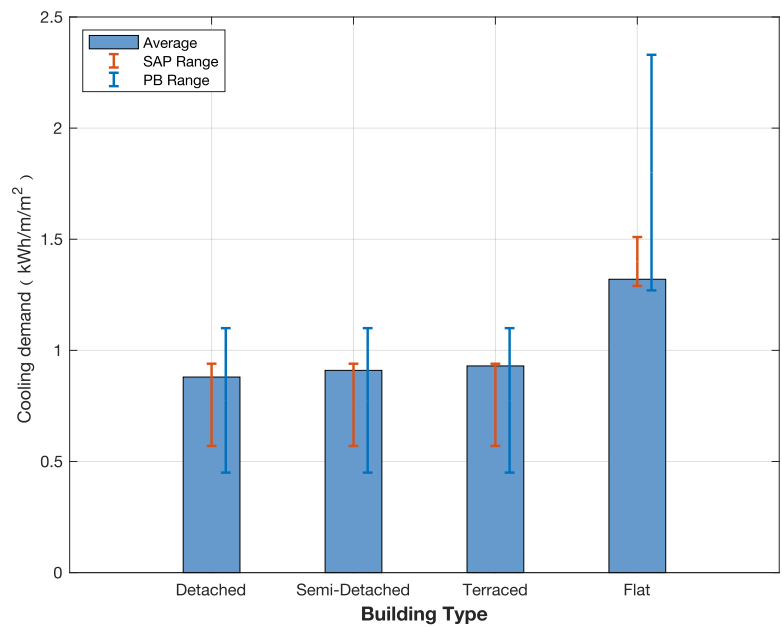

Figure 3 Comparison of average cooling demand with SAP and PB ranges

2.3 Time series prediction model

RNNs are a class of neural networks designed to process sequential data by capturing temporal dependencies. However, standard RNNs often encounter issues like vanishing or exploding gradients when handling long sequences, which limits their effectiveness in practical applications. To overcome these challenges, two advanced RNN architectures—LSTM and GRU—were introduced [38] [39]. These models enhance the ability to learn long-term dependencies through specialized mechanisms. LSTM employs memory cells equipped with three gates—input, forget, and output gates—that regulate the flow of information. This architecture enables LSTM to selectively retain relevant information while discarding unnecessary data, effectively addressing gradient-related issues. As a result, LSTM is well-suited for tasks involving long-term sequential patterns, such as time series forecasting. GRU, in contrast, offers a simpler alternative by combining the functionality of LSTM's memory cells and hidden states into a unified structure. It uses update and reset gates to manage information flow, maintaining efficiency without compromising performance. GRU's compact architecture often makes it a favourable choice in scenarios with limited computational resources. More details about the structure and mathematics of LSTM and GRU can be found in [40].

2.4 Training data for the model

The simulation input data for all buildings is highly complex, making it impractical to include every building's detail in the LSTM or GRU model for cooling prediction across London. Instead, LSTM and GRU simplify the process by using only meteorological data and time of day as inputs, predicting London's total cooling demand as the output. The input meteorological data includes dry-bulb temperature, wet-bulb temperature, atmospheric pressure, relative humidity, global horizontal irradiance (GHI), direct normal irradiance (DNI) and diffuse horizontal irradiance (DHI), all of which capture the environmental conditions affecting heating demand. The time data provides temporal context and includes hour of the day, allowing the model to learn patterns based on daily variations. The historical weather data from the summer of 2020 in London, along with the corresponding time-related data, was selected as the simulation year for the model. The dataset was split into two parts: the 80% of the weather and time-related data, along with results from the DB simulations, was used for training, while the remaining 20% was reserved for testing the model's performance.

2.5 Bayesian Optimisation (BO)

The BO process starts with a few initial random evaluations of different hyperparameter settings to create a base for the surrogate model. Once this base is set, the surrogate model (like Gaussian Process) is trained to estimate the objective function. The acquisition function then chooses the next best hyperparameter setting to test. After each evaluation, the new data is added to improve the surrogate model. This loop continues, with each step refining the model and narrowing the search. In this study, BO is used to tune hyperparameters for LSTM and GRU models, such as learning rate, hidden layer size, and batch size. By guiding the search in a systematic way, BO finds near-optimal configurations with fewer tests, enhancing both model performance and efficiency, making it a crucial tool for tuning complex neural networks.

2.6 Data partitioning methods

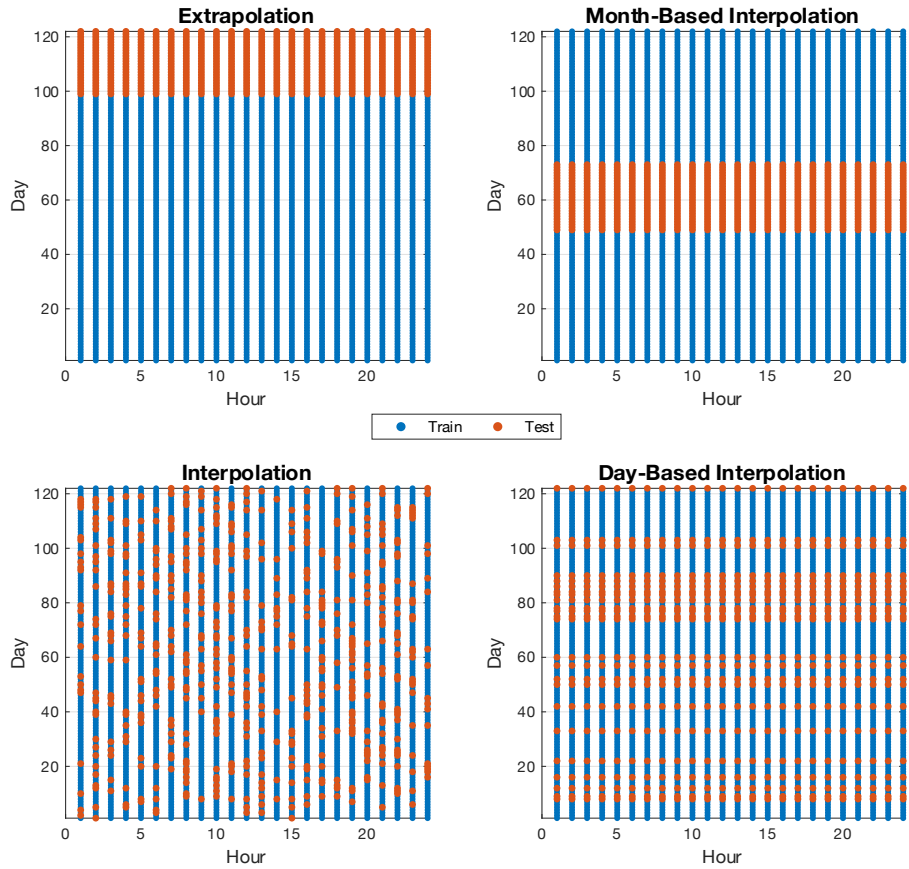

**Figure 4 Data partitioning methods: extrapolation, month-based interpolation, interpolation and day-based interpolation**

As shown in Figure 4, four data partitioning methods are used in this study. Overall, extrapolation refers to making predictions beyond the known data range, where the test data falls outside the distribution of the training data. Interpolation refers to making predictions within the known data range, where the test data is within the distribution of the training data. Specifically, extrapolation in this study uses the first 80% of the dataset as training data and the last 20% as testing data, maintaining the original temporal order. The second method, referred to as month-based interpolation, selects a continuous 20% dataset as the test set, while the remaining 80% is reordered chronologically and used for training. As shown in Figure 4, in a 122-day dataset, data from Day 44 to Day 68 would be selected as the test set, while the data from Day 1 to Day 43 and Day 69 to Day 122 would be used for training. The third method is the most common form of interpolation, where 20% of the data is randomly selected as the test set, while the remaining 80% is used for training. In this case, the time series is shuffled, and the data is no longer arranged in the temporal order of a 24-hour cycle or the sequential order of days from 1 to 122. The final method, referred to as day-based interpolation, involves randomly selecting data from 20% of the days as the test set while preserving the temporal structure within each day. It can also be seen from Figure 4 that, compared to interpolation, day-based interpolation preserves the temporal sequence within each day in the training and testing data, but the order of days from 1 to 122 is shuffled.

## 3. Results and analysis

This section begins with an analysis of the BO process and its results. Then, stepwise training is conducted to explore the impact of the optimisation process on model performance. It shows improvements when either model parameters or training parameters are optimised individually. Further enhancements are observed when both are optimised together, validating the effectiveness of BO. The data partitioning method in Section 3.1 and Section 3.2 is extrapolation. In section 3.3, the performance of GRU and LSTM is compared under extrapolation, month-based interpolation, interpolation and day-based interpolation modes. All types of data partitioning models have been optimised using BO. Also, the prediction results of the optimised model are evaluated. Finally, different types of models are used to predict cooling demand in the summer of 2050 to assess their generalisation ability.

3.1 Analysis and results of BO

Table 1 lists the hyperparameters used in BO along with their respective ranges, divided into two main categories: model structure parameters and training process parameters. Model structure parameters and training process parameters differ in their roles and impact. Model structure parameters define the architecture of the model, such as the number of layers, the number of neurons per layer, and the activation functions. These parameters determine how the model processes data and extracts patterns. In contrast, training process parameters control how the model is trained, including factors like learning rate, batch size, and the number of training epochs. The model structure parameters include settings that define the depth and size of the network. Number of layers, set between 1 and 5, controls the depth of the network by determining how many hidden layers are included. Increasing the number of layers allows the model to learn more complex, hierarchical patterns, but it also adds to the computational requirements and the risk of overfitting. Number of hidden units in Layer 1 through Layer 5 each range from 10 to 200, specifying the number of neurons in each layer. A higher number of units per layer enhances the model's ability to capture complex patterns but also increases computational costs.

Table 1 Hyperparameters for BO in model structure and training process

| Parameter | Minimum value | Maximum value | Optimised value |
|---|---|---|---|
| **Model hyperparameter optimisation** | | | |
| Number of hidden Layers | 1 | 5 | 5 |
| Neurons in each layer | 10 | 200 | [112, 81,178, 56, 175] |
| **Training hyperparameter optimisation** | | | |
| Initial learning rate ($\eta_0$) | 0.0001 | 0.01 | 0.0069 |
| Learning rate drop period ($P$) | 50 | 200 | 131 |
| Learning rate drop factor ($F$) | 0.1 | 0.5 | 0.1194 |
| Max epochs | 50 | 200 | 183 |
| Mini batch size | 16 | 128 | 27 |
| Gradient threshold | 0.5 | 2 | 1.3198 |

The remaining parameters fall under training process parameters, which influence how the model learns over time. Initial learning rate, set within the range of 0.0001 to 0.01, determines the step size for each parameter update, impacting both the speed and stability of convergence. The learning rate drop period, ranging from 50 to 200, controls the frequency (in epochs) at which the learning rate is reduced, enabling the model to gradually refine its learning as it progresses. Learning rate drop factor, with a range of 0.1 to 0.5, specifies the factor by which the learning rate is multiplied at each drop period, facilitating a gradual reduction in learning rate to improve convergence as the model approaches optimal values. For the piecewise learning rate schedule in the model, the formula for learning rate is:8

$$\eta_k = \eta_0 \cdot F^n, \text{ where } n = \lfloor \frac{k}{P} \rfloor$$

$\eta_k$ is learning rate at epoch $k$; $\eta_0$ is initial learning rate; $F$ is learning rate drop factor; $P$ is learning rate drop period (in epochs); $n$ is number of completed learning rate decay steps, calculated as $n = \lfloor \frac{k}{P} \rfloor$

(floor function), which takes the integer part of $\frac{k}{P}$. Max epochs, ranging from 50 to 200, specifies the maximum number of complete passes through the training data, allowing more training iterations to refine the model but also increasing computational cost. Mini batch size is defined between 16 and 128, indicating the number of samples processed in each training batch. Larger batch sizes can provide more stable training results, although they require additional memory and processing power. Lastly, the gradient threshold, ranging from 0.5 to 2, helps prevent gradient explosion during backpropagation by limiting the gradient values, thereby stabilizing the training process. Through BO, these parameters are intelligently explored within their defined ranges to identify the optimal configuration, ultimately enhancing model performance while managing computational costs.

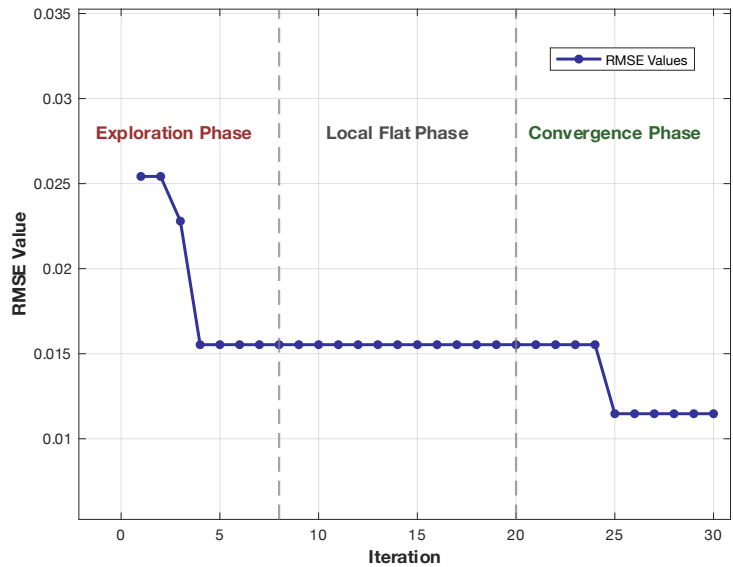

Figure 5 RMSE optimisation progression during Bayesian hyperparameter tuning

Figure 5 shows the root mean square error (RMSE) progression during the BO process over 30 iterations. All variables are integer with a fixed step size of 1, except real-valued variables (e.g., gradient threshold), which are dynamically sampled, and log-transformed variables (e.g., initial learning rate), which are sampled at equidistant intervals on a logarithmic scale with exponentially varying values. Initially, the RMSE decreases rapidly from approximately 0.026 to 0.016 within the first few iterations, reflecting the optimisation algorithm's ability to quickly identify promising parameter configurations. Between iterations 5 and 24, the RMSE stabilizes, indicating a plateau in performance as the algorithm explores less impactful adjustments. Around iteration 25, a further drop in RMSE to approximately 0.012 is observed, suggesting that the optimisation process identified a significantly better set of hyperparameters. The smoothed curve effectively highlights these phases, showing the rapid initial improvement, the exploration phase, and the final refinement, demonstrating the efficiency of BO in enhancing model performance. Correspondingly, during the RMSE optimisation process, the optimised model parameters and training hyperparameters were identified as shown in Table 2. These include a model configuration with 5 layers and hidden units set to 112, 81, 178, 56, and 175, respectively. Training hyperparameters were optimised to a learning rate of 0.0069, a mini-batch size of 27, 183 maximum epochs, and a learning rate drop factor of 0.1194 applied every 131 epochs. These configurations played a crucial role in achieving the minimized RMSE observed during the optimisation process.

3.2 Results for different optimisation level

Table 2 presents a detailed breakdown of the performance of LSTM and GRU models across different stages of optimisation: the original model, model hyperparameter optimisation, training hyperparameter optimisation, and combined model and training hyperparameter optimisation. Each stage records the

RMSE and mean absolute error (MAE) on both the training and test sets, allowing for an in-depth evaluation of how these optimisation strategies affect model accuracy and generalization. In the original model, both LSTM and GRU exhibit relatively high error rates on the training and test sets, with only moderate generalization. Specifically, for LSTM, the training RMSE is 2.04% with an MAE of 0.69%, and the test RMSE is 2.27% with an MAE of 0.91%. GRU follows a similar pattern, with a training RMSE of 2.16% and MAE of 0.73%, and a higher test RMSE of 2.32% with MAE reaching 1.40%. These initial results highlight that both models, while functional, could benefit from further fine-tuning, as indicated by the slightly higher errors on the test set compared to the training set.

**Table 2 Performance metrics of LSTM and GRU models across different optimisation stages**

| Model | Dataset | RMSE (%) | MAE (%) |
|---|---|---|---|
| **Original** | | | |
| LSTM | Training | 2.04% | 0.69% |
| LSTM | Test | 2.27% | 0.91% |
| GRU | Training | 2.16% | 0.73% |
| GRU | Test | 2.32% | 1.40% |
| **Model hyperparameter optimisation** | | | |
| LSTM | Training | 5.06% | 2.22% |
| LSTM | Test | 2.00% | 0.97% |
| GRU | Training | 4.17% | 1.86% |
| GRU | Test | 1.91% | 1.10% |
| **Training hyperparameter optimisation** | | | |
| LSTM | Training | 2.98% | 1.16% |
| LSTM | Test | 1.79% | 0.72% |
| GRU | Training | 1.45% | 0.72% |
| GRU | Test | 1.98% | 0.68% |
| **Model and training hyperparameter optimisation** | | | |
| LSTM | Training | 1.91% | 0.85% |
| LSTM | Test | 1.50% | 0.85% |
| GRU | Training | 2.12% | 0.86% |
| GRU | Test | 1.32% | 0.55% |

After optimising only, the model hyperparameters, an increase in training set errors is observed for both models, suggesting a shift in the balance toward better test performance, even at the cost of slightly overfitting the training data. The LSTM's test RMSE decreases from 2.27% to 2.00%, and MAE from 0.91% to 0.97%. Similarly, GRU's test RMSE drops to 1.91% and MAE to 1.10%. Despite the increase in training error (e.g., LSTM's training RMSE rises to 5.06% and MAE to 2.22%), the overall reduction in test error implies improved generalization. This suggests that optimising structural hyperparameters (such as the number of hidden units or layers) can help models handle unseen data better, although the increase in training error points to the need for further tuning. When focusing solely on training hyperparameter optimisation, the results indicate enhanced accuracy across both training and test sets. The LSTM model's test RMSE further decreases to 1.79%, with MAE dropping to 0.72%, showing a clear improvement over the original model and the model-only optimisation. GRU's test performance also improves, with RMSE dropping to 1.98% and MAE to 0.68%. Notably, the training error for GRU is substantially reduced, with an RMSE of 1.45% and an MAE of 0.72%, indicating a more efficient learning process. This stage demonstrates that training parameters (such as learning rate, batch size, and epochs) significantly influence the model's convergence and ability to generalize. Lower test errors across both models show that training hyperparameter tuning effectively mitigates overfitting, balancing improvements on both the training and test sets.

The final stage, which combines both model and training hyperparameter optimisation, yields the best performance, with notable improvements on both the training and test sets. For LSTM, the test RMSE reaches its lowest value of 1.50%, and the MAE is reduced to 0.85%. GRU shows even greater improvements, achieving a test RMSE of 1.32% and a significantly reduced MAE of 0.55%. In the training set, both LSTM and GRU maintain competitive performance, with GRU showing slightly higher training error than LSTM (training RMSE of 2.12% for GRU vs. 1.91% for LSTM). This balanced reduction in training and test errors highlights the benefit of a holistic approach to optimisation, where jointly adjusting both model and training hyperparameters allows the models to achieve the highest accuracy while minimizing overfitting. Interestingly, GRU generally exhibits lower test errors than LSTM at each stage, suggesting that GRU may be better suited for capturing patterns in time-series data, such as cooling demand forecasting, under the specified configurations.

3.3 Results comparison of interpolation and extrapolation

Table 3 illustrates the performance comparison between GRU and LSTM on the training and testing dataset, evaluated through regression analysis of the prediction and actual values. Additional regression plots for the training sets are provided in Appendix 1, offering a clearer view of the models' behaviour during training process. Both GRU and LSTM models exhibit a strong linear relationship between predicted and actual values, indicating high predictive accuracy in the extrapolation task. In terms of numerical results, GRU achieves an $R^2$ of 0.9898, slightly higher than LSTM's $R^2$ of 0.9589, suggesting that GRU has a slight edge in extrapolation performance. Overall, both GRU and LSTM exhibit excellent predictive capabilities, accurately capturing the extrapolated trends even beyond the training data range. In comparison, the training results of month-based interpolation are worse than those of extrapolation, especially for LSTM. With an $R^2$ of 0.9571 compared to LSTM's 0.8721, GRU demonstrates superior fitting performance on the training set. In interpolation, GRU and LSTM exhibit a certain degree of linearity between predicted and actual values, but this linear relationship is weaker compared to their performance in extrapolation tasks. LSTM outperforms GRU, achieving an $R^2$ of 0.7962 compared to GRU's $R^2$ of 0.7225, indicating that LSTM provides higher prediction accuracy within the known range. For day-based interpolation model, the $R^2$ values for GRU is 0.9489 and LSTM is 0.895, indicating that this method that preserves the time series structure is more suitable for this model. The test results are also compared to further analyse the impact of the three data splitting methods on the model.

Table 3 Regression results of test set

| Train | Extrapolation | Month-based interpolation | Interpolation | Day-based interpolation |
|---|---|---|---|---|
| LSTM | 0.9589 | 0.8721 | 0.7962 | 0.8955 |
| GRU | 0.9898 | 0.9517 | 0.7225 | 0.9489 |
| Test | Extrapolation | Month-based interpolation | Interpolation | Day-based interpolation |
| LSTM | 0.7114 | 0.7282 | 0.7890 | 0.9108 |
| GRU | 0.8595 | 0.7659 | 0.7596 | 0.9386 |

Compared with training set, $R^2$ for both extrapolation models decrease. It indicates that both models lack sufficient generalization ability. GRU's $R^2$ for extrapolation decreases from 0.9898 on the training set to 0.8595 on the test set, while LSTM's $R^2$ for extrapolation drops from 0.9587 to 0.7114. For month-based interpolation, $R^2$ for GRU is 0.7659 and $R^2$ for LSTM is 0.7282. This difference arises primarily because the test set in the extrapolation task extends beyond the training data range, forcing the models to rely solely on global trends for predictions. In contrast, the interpolation task shows a smaller gap between training and test set $R^2$. GRU achieves $R^2$ of 0.7225 on the training set and 0.7596 on the test set, while LSTM achieves 0.7962 and 0.7890, respectively. This is because the test set in the interpolation task shares a similar distribution with the training set, and the models do not need to handle unseen regions. Their performance depends on their ability to fit the local data distribution, which remains consistent between training and test sets. The large difference between training and test set $R^2$ in the extrapolation task reflects the models' uncertainty in handling unseen regions, whereas the smaller difference in the interpolation task indicates stable generalization within the known range. The $R^2$ for GRU and LSTM under the day-based interpolation method are 0.9386 and 0.9108, respectively, which are on the same level as the training results of 0.9489 and 0.8955. Day-based interpolation combines the advantages of both interpolation and extrapolation. It preserves the temporal order within each day while also ensuring sufficient randomness in the selection of training data.

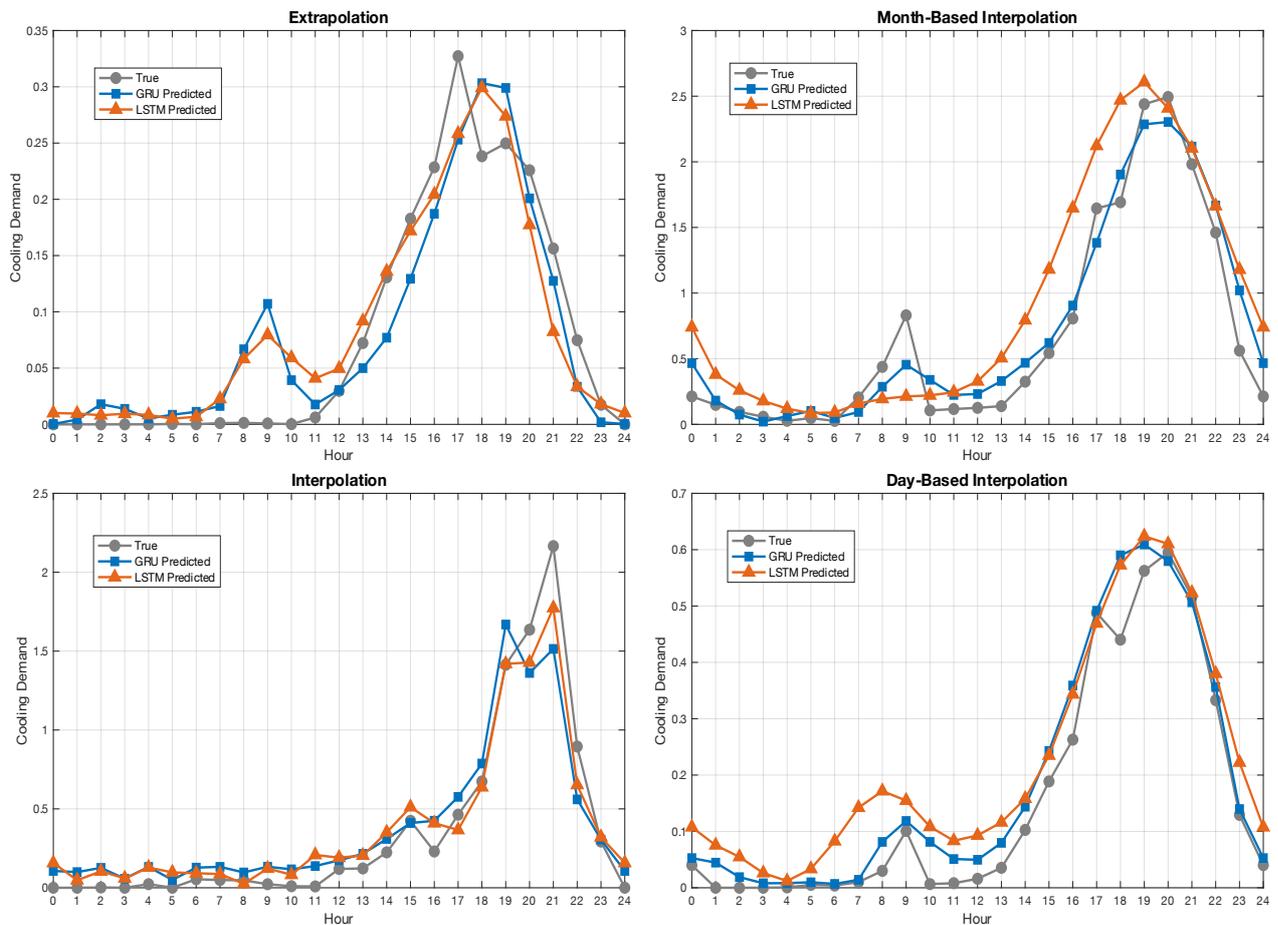

**Figure 6 Average hourly cooling demand prediction: extrapolation, month-based interpolation, interpolation and day-based interpolation for test sets (GRU vs LSTM)**

Figure 6 compares the impact of four different test set partitioning strategies, including extrapolation, month-based interpolation, interpolation, and day-based interpolation, on the performance of GRU and LSTM models in predicting cooling demand. The grey line represents the true cooling demand, while the blue and orange lines correspond to the GRU and LSTM predictions, respectively. In the extrapolation method, the cooling demand ranges approximately from 0 to 0.35, reflecting an overall low cooling demand. The true values exhibit a single-peak trend, with the main cooling demand concentrated in the afternoon hours. However, both predictions of LSTM and GRU show a small peak between 8 a.m. and 10 a.m.. This is because the test data is entirely taken from the last 20% of the time series, which was completely unseen during training. The training data is significantly larger than the test data, and a fluctuation exist in the training data between 8 a.m. and 10 a.m., leading to higher model predictions during this period. In the month-based interpolation method, the cooling demand ranges from 0 to over 2.5, and the true values exhibit a double-peak trend, with peaks in both the morning and the afternoon. The morning peak is captured by GRU but missed by LSTM. While both models can detect the afternoon peak, LSTM's error is noticeably larger than GRU's. The results of this sampling method largely depend on the differences between the training and test datasets, which is similar to extrapolation. If the characteristics of the training data do not adequately cover those of the test data, it can lead to inaccurate predictions. Interpolation does not have such concerns, as the training and test data are randomly selected. The data distribution remains balanced, reducing the likelihood of extreme cases. Although the predicted values tend to be higher when the demand is below 0.5 and lower when the demand exceeds 1.5, both GRU and LSTM accurately capture the overall trend of cooling demand. This discrepancy could be caused by the disruption of the time series, preventing the models from learning the hidden cooling schedule embedded in the sequential data. In the day-based interpolation method, the cooling demand ranges from 0 to 0.7, and the plot shows that it preserves the daily time sequence. This profile is similar to extrapolation and month-based interpolation, but it performs better because both LSTM and GRU successfully capture the two peaks in cooling demand.

3.4 Results of optimal model

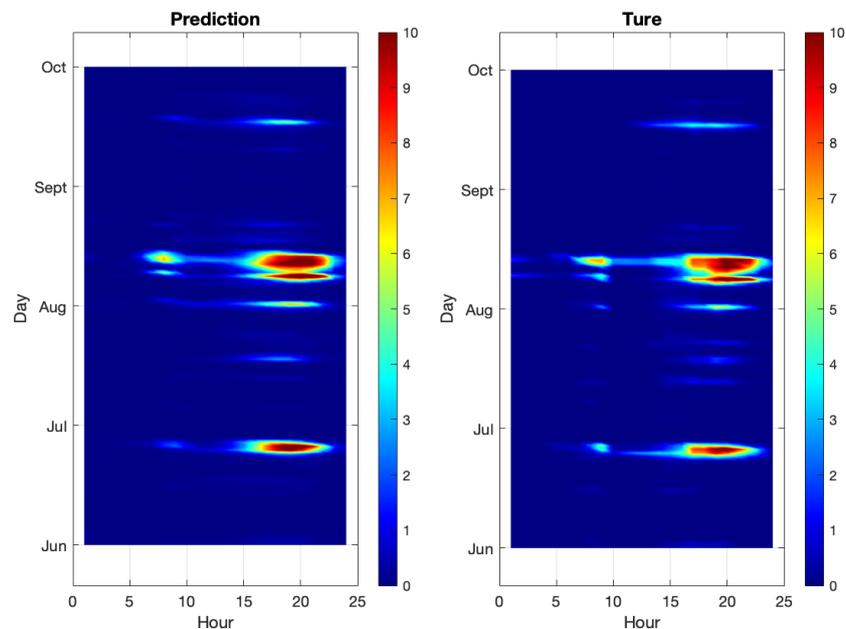

**Figure 7 Temporal cooling demand heatmap of optimal model**

Among all model results, the GRU with day-based interpolation demonstrated the best performance across different evaluation metrics. This model achieved the highest $R^2$ (0.9386), indicating the best overall fit to the true values. Fig. 7 presents a contour plot comparison between the predicted and actual cooling demand of this model. The x-axis represents the hours of the day (0–24 hours), while the y-axis corresponds to the days. The colour intensity indicates the cooling demand, with warmer colours (yellow to red) signifying higher demand and cooler colours (blue) representing lower demand. The cooling demand is highest during June, July, and August, particularly in the afternoon, which aligns with the typical summer cooling demand pattern. In September, the cooling demand decreases, indicating a reduction in cooling needs as temperatures drop. The model accurately captures the major peak regions from June to August, especially in the afternoon, closely matching the trends observed in the true values. This suggests that the model not only successfully identifies cooling demand peaks and overall seasonal trends but also maintains high predictive accuracy across different time periods.

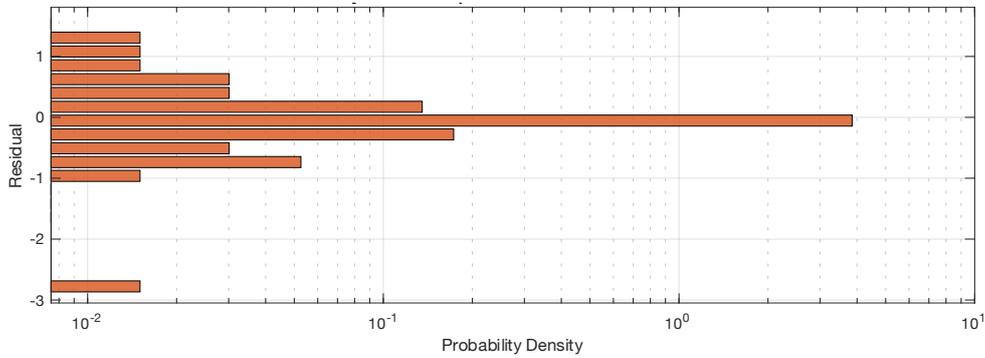

**Figure 8 Residual distribution for day-based interpolation GRU predictions on testing datasets**

The histogram in Fig. 8 presents the residual distribution for day-based interpolation in terms of probability density. Residuals represent the difference between the true and predicted values. The y-axis denotes the residual values (prediction errors), while the x-axis, shown on a logarithmic scale, represents the probability density of each residual value. The distribution appears slightly skewed with a higher occurrence of negative residuals, implying that the model tends to overestimate demand more frequently in this setting. Residuals are primarily concentrated around 0, ranging from -1 to 1, indicating that the model's prediction errors are generally small and well-controlled. However, extreme residuals, such as -3, are also present. It is important to note that these extreme residuals are rare, as reflected by the low probability density on the logarithmic scale. These outliers account for only a small fraction of the data and do not significantly affect the overall distribution. The majority of residuals in day-based interpolation remain clustered around 0, reinforcing the model's overall reliability in prediction.

3.5 Prediction for 2050

**Table 5 Performance of LSTM and GRU models in prediction of 2050 cooling demand**

| GRU | RMSE | MAE |
| --- | --- | --- |
| Extrapolation | 5.411% | 1.988% |
| Interpolation | 6.556% | 3.559% |
| Day-based interpolation | 6.027% | 2.054% |
| Month-based interpolation | 6.881% | 2.446% |
| LSTM | RMSE | MAE |
| Extrapolation | 5.984% | 2.088% |
| Interpolation | 15.679% | 6.923% |
| Day-based interpolation | 6.107% | 2.309% |
| Month-based interpolation | 5.706% | 2.655% |

This study evaluated the performance of these models in predicting cooling demand for 122 days in the summer of 2050, with the results presented in Table 5. A comparison with Table 2 reveals that the test results for 2020 are generally better than the predictions for 2050. Since the prediction period for 2050 exceeds the time span of the training data, and the weather conditions in 2050 differ significantly from those in 2020, the models must make predictions under new climate conditions, posing a greater challenge to their generalisation ability. The GRU models exhibit more stable performance, particularly in interpolation, where their RMSE (6.556%) and MAE (3.559%) are significantly lower than those of LSTM (RMSE of 15.679% and MAE of 6.923%). This suggests that GRU has a stronger ability to capture local patterns in time series and can better handle missing values in the predictions. In contrast, LSTM demonstrates higher errors in interpolation, possibly due to the disruption of temporal order in this task, which limits the effectiveness of its memory units. In other scenarios, the performance gap between the two models is smaller. For instance, in extrapolation, GRU and LSTM show similar results, with GRU achieving an RMSE of 5.411% and MAE of 1.988%, slightly better than LSTM (RMSE of 5.984% and MAE of 2.088%). This indicates that both GRU and LSTM are capable of capturing long- and short-term dependencies in continuous time series forecasting. Additionally, in day-based interpolation, GRU achieves an RMSE of 6.027% and MAE of 2.054%, which is also slightly better than LSTM (RMSE of 6.107% and MAE of 2.309%). However, in month-based interpolation, LSTM performs slightly better, with an RMSE of 5.706% and MAE of 2.655%, suggesting that it has a certain advantage in handling long-term forecasting tasks.

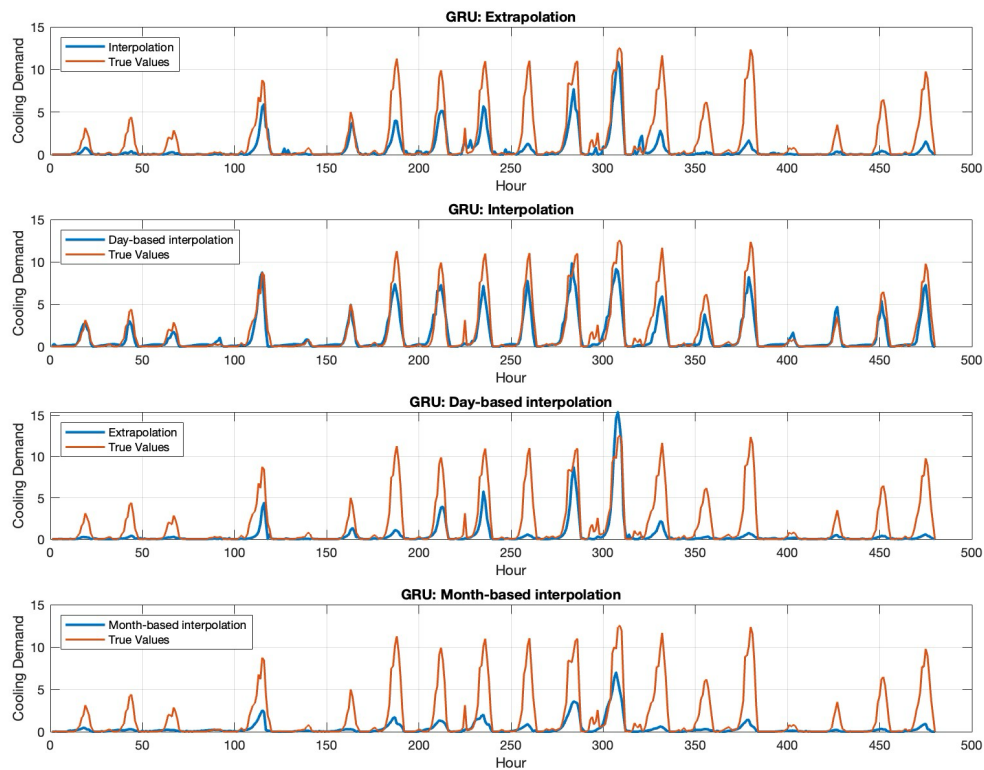

Figure 9 The performance of different GRU models in predicting peak cooling demand in 2050

Figure 9 illustrates the performance of different GRU models in predicting peak cooling demand. The differences are quite significant, with interpolation performing the best. During peak periods, its predictions closely match the actual values, indicating that GRU with interpolation can effectively capture local patterns in the time series and the fluctuations in peak demand. In comparison, the performance of extrapolation and day-based interpolation is slightly inferior to interpolation. Some peak values are underestimated, yet the model is still able to capture certain peak demand patterns, demonstrating that GRU has a strong ability to identify trends in extrapolation and short-term interpolation tasks. In month-based interpolation, GRU tends to underestimate peak demand, especially around significant peak values. This suggests that its ability to capture local patterns diminishes when dealing with long-term interpolation tasks. Overall, GRU performs best in interpolation, while its performance in extrapolation and day-based interpolation remains relatively stable.

## 4. Conclusion

In this study, a cooling demand prediction framework combining physical modelling and deep learning was developed to address the challenges posed by limited historical data and the absence of high-resolution cooling demand prediction models. Four distinct partitioning strategies were implemented to improve model performance under limited data scenarios. Day-based interpolation demonstrated clear advantages for cooling demand prediction among four data partitioning methods.

Both the GRU and LSTM models showed significant improvements after Bayesian optimisation, achieving better accuracy and generalisation. By preserving the daily time sequence while incorporating randomness in sample selection, the GRU model with day-based interpolation effectively captures both local variations and overall trends. When forecasting cooling demand for future climate scenarios in 2050, under out-of-domain climate conditions, the global interpolation method proved more effective in capturing long-term trends and peak demands. This highlights the importance of selecting suitable data partitioning strategies to ensure model generalisation when applied to significantly different environmental conditions.

In future work, several promising directions can be explored to further enhance the model ability. Embedding the underlying physical principles to develop physics-enhanced neural networks offers a way to improve accuracy and robustness. Additionally, developing adaptive data partitioning strategies that dynamically adjust to data characteristics could further enhance model generalisation. Combining above strategies with transfer learning would allow models trained under current climate conditions to better adapt to future climates.

**Acknowledgement**

The authors like to acknowledge EPSRC in the UK for financial support through the grants EP/X025322/1, EP/T022701/1, EP/V042033/2, EP/V030515/1, and EP/W027593/2, and the Royal Society for their financial support (Ref: IF\R1\231053). This project has also received funding from the European Union's Horizon 2020 research and innovation programme under the Marie Sklodowska-Curie grant agreement No. 101007976.

# Appendix 1

This appendix presents the regression analysis plots for different data partitioning strategies, including extrapolation, month-based interpolation, global interpolation, and day-based interpolation. These plots compare the performance of LSTM and GRU models on the training set under each partitioning strategy.

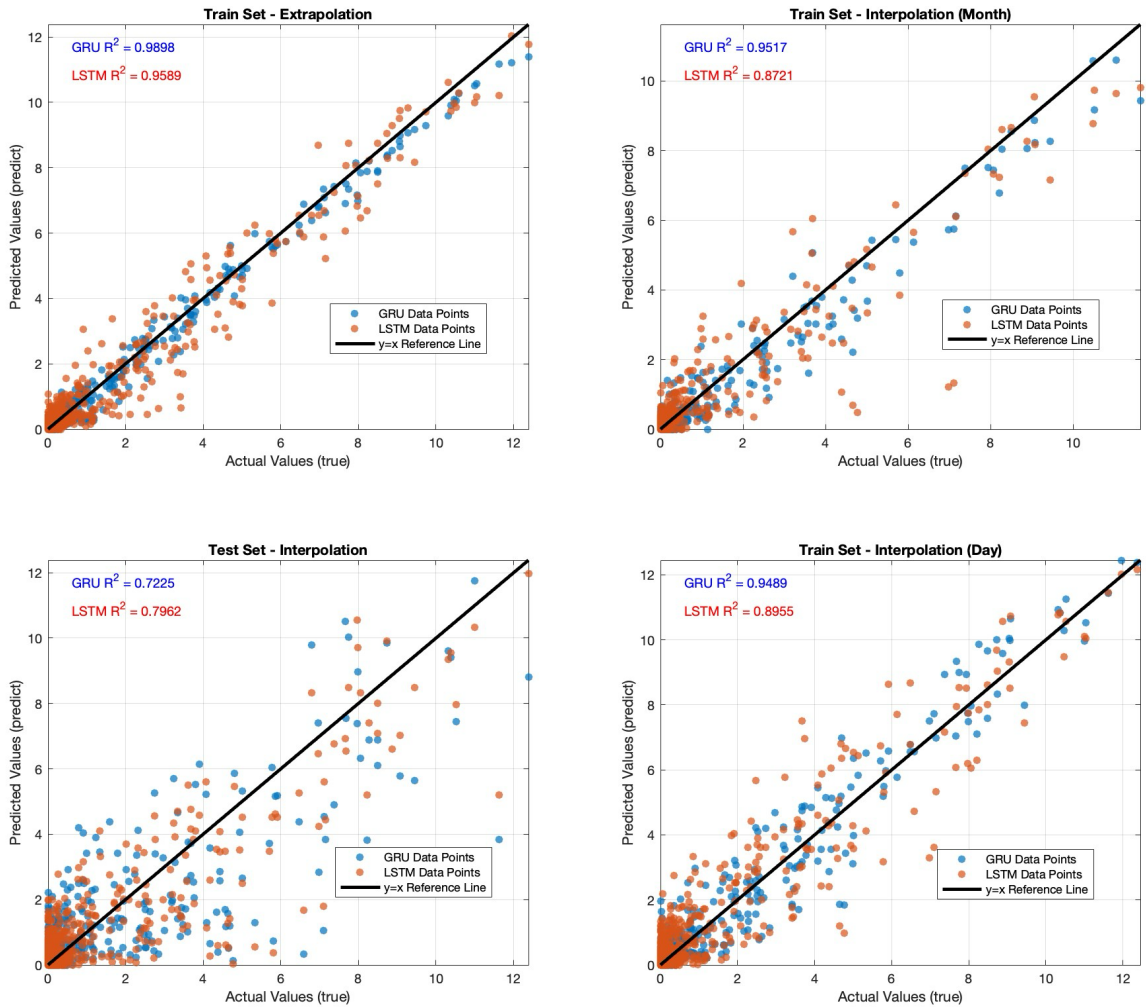

**Fig. A1 Regression analysis on training dataset: extrapolation; month-based interpolation; interpolation and day-based interpolation (LSTM vs GRU)**